# Electrons and phonons in pentacene: coupling patterns reveal the microscopic origin of the phonon limited mobility

Luca Gnoli[1,]*, Elisabetta Venuti[2], Tommaso Salzillo[2], Matteo Masino[3], Patrizio Graziosi[1]

[1] ISMN – CNR, Consiglio Nazionale delle Ricerche, via Gobetti 101, 40129, Bologna, Italy

[2] Dipartimento di Chimica Industriale "Toso Montanari", Università di Bologna, via Gobetti 85, 410129, Italy

[3] Dipartimento di Scienze Chimiche, Della Vita e Della Sostenibilità Ambientale & INSTM-UdR Parma, Parco Area delle Scienze, 17/A, 43124 Parma, Italy

* lucagnoli@cnr.it

## Abstract

We have performed a comprehensive computational study of the vibrational properties and electron–phonon couplings in the three known polymorphs of pentacene. Vibrational patterns and electron–phonon interactions were calculated at several $\boldsymbol{q}$-points of the Brillouin zone, allowing for a detailed mapping of the phonon landscape and the associated coupling mechanisms relevant to charge transport. Using a pool of post-processing tools, we analyze the different phonon dispersions. Thus, we shed light on how low-frequency phonons modulate the transport differently in the polymorphs via their distinct electron-phonon coupling (EPC) signatures in reciprocal space. In fact, we show that distinct phonons dominate in high-temperature and thin-film polymorphs with respect to the high mobility low-temperature polymorph, and that these lead to different decoherence/localization trends. We describe the microscopic origin of the mobility in bulk polymorph, demonstrating that polymorphism not only modulates the transfer integral but also the phonon pattern. For the first time we show that different EPC patterns lead to very different mobility values even in similar structures. Also, we explain how phonon confinement is responsible for the increased mobility observed in 2D phase.

Finally, we address the problem of possible coexistence of multiple polymorphs within a single specimen, frequently encountered in organic crystals due to their subtle energy landscape and processing conditions. In this context, we consider the implications of polymorph intergrowth, structural defects and disorder.

## I. Introduction

Pentacene has been for a long time regarded as a benchmark for Organic Semiconductors (OSCs). Although in recent years research on OSCs has advanced towards novel, functionalized systems, pentacene remains an ideal reference system for developing fundamental knowledge, as it is considered a model p-type semiconductor.[1–4] This is of special importance because the widespread application of OSCs in electronics-related applications is hampered by the lack of predictive understanding of the inter-relationship between solid-state packing and device performance, irrespective of the specific type of device. This challenge also arises from the likely occurrence of polymorphism, where a single molecular compound can crystallize in multiple distinct structures, each potentially exhibiting different electronic and vibrational properties. Thus, each polymorph can exhibit distinct variations in the electron-phonon coupling (EPC), thereby influencing the transport characteristics of the OSC.[5,6] Indeed, low-frequency vibrational modes, such as translations and rotations of the entire molecule, play a pivotal role in introducing dynamic disorder which, in turn, affects the electronic transfer integrals and hence the bandwidth, ultimately shaping the macroscale electronic properties of the semiconductor. Therefore, understanding polymorphism is crucial for defining the electronic properties in OSCs at the macroscopic scale.

In this study, we first present accurate and validated first-principles calculations of the three known pentacene polymorphs: low temperature (LT), high temperature (HT) and thin film (TF). We use the inherent structures method [7–10] to ensure they are really three different structures, and assess their vibrational fingerprints. Then, we compute the EPC at the zone-boundary and demonstrate that the strongest EPC happens for modes having non-zero momentum, i.e. not in $\Gamma$. We can then assign the experimentally identified modes – a task still missing so far. Finally, we use the computed EPC to evaluate the mobility $\mu$ [11] of the three polymorph and related mixture.[12] Also, we use pentacene as a benchmark to test the computational description of the role of disorder or defects in terms of energy potential and carrier scattering.

## II. Computational method

### a. Electronic and Vibrational properties

We employ PBE-PAW pseudopotentials and the D3-BJ Grimme with Becke-Johnson damping functions for a posteriori VdW correction *at any stage* of a DFT calculation with VASP. [11,13–15] We start from the experimental unit cell parameters and determine the optimal ***k***-points grid using a cutoff energy of 400 eV. After identifying a converging ***k***-grid, we optimize the energy cutoff for the wavefunctions. Finally, we relax the atomic positions keeping constant the unit cell parameters at their experimental values. For the study of inherent structures, we relax also the lattice parameters. Following relaxation, the Phonopy package [16,17] is used to compute and diagonalize the dynamical matrix using a 2×2×2 super-cell – a setup validated to ensure convergence in phonon frequency and 3D density of states (DOS) with no negative frequency, except for the acoustic branches in Γ, where small negative frequency around $10^{-1}$ - $10^{-2}$ THz may appear and considered acceptable. Finally, the off-resonant Raman activities are computed with the vasp_raman.py program.[18] The program uses VASP as back-end to compute the polarizability with the finite displacements approach, and returns the Raman activity of the selected modes.

Throughout the text, we use $\boldsymbol{k}$ and $\boldsymbol{q}$ to indicate the BZ points, and hence their wave-vector, of electrons and phonons, respectively.

### b. Electron-phonon coupling (EPC)

We adopt a recently developed protocol to parametrize the EPC. [11] The unit cell is modulated along the selected eigenmodes at the Γ point and all the relevant high-symmetry $\boldsymbol{q}$-points within the Brillouin Zone (BZ), as defined in the Bilbao Crystallographic Data Center. [19,20] Therefore, a unit cell distorted along the eigenmode vibrational coordinate is generated for each considered $\boldsymbol{q}$ point and $\nu$ phonon branch. The Phonopy code automatically accounts for the atomic mass weights in this process. Considering that in the tight-binding approximation the electronic energy dispersion along a 1D direction can be approximated to $\sim 2\, t_W\, cos(\boldsymbol{k} \cdot \boldsymbol{a})$, that leads to a bandwidth $B_W$ related to the transfer integral $t_W$ by the relation $B_W = 4t_W$. [21] As a result, the comprehensive bandwidth across the full 3D BZ, and hence the corresponding transfer integral, can be obtained from a first principle DFT electronic structure calculation.

Under this conceptual framework, and taking into account both the traditional approach to EPC [22,23] and the conventional definition of Deformation Potentials, [24,25] the EPC constant for each $\boldsymbol{q}$-point and phonon branch $\nu$ can be defined as:[11]

$$D_{\boldsymbol{q},\nu}^{n,n} = \frac{\partial t_W^n}{\partial \boldsymbol{r}_{\boldsymbol{q},\nu}} \quad (1)$$

$$D_{\boldsymbol{q},\nu}^{n,m} = \frac{\partial \Delta_s^{n,m}}{\partial \boldsymbol{r}_{\boldsymbol{q},\nu}} \quad (2)$$

In Eqs. (1) and (2), $n$ and $m$ are the band indexes, $t_W^n$ is the bandwidth of the band of index $n$, $\Delta_s^{n,m}$ is the Davydov splitting between the bands *n* and *m*, computed as the energy difference between the energy barycenters of the bands, $\boldsymbol{r}_{\boldsymbol{q},\nu}$ is the average displacements of all the atoms in the structure distorted along the eigenmode $(\boldsymbol{q}, \nu)$. Thus, Eq. (1) is related to *intra*-band processes, i.e. will be used when the carrier scattering involves initial and final states in the same band, while Eq. (2) is related to *inter*-band processes.

Operatively, after a self-consistent DFT calculation for each $(\boldsymbol{q}, \nu)$ modulated/eigen-distorted structure, we perform a non-self-consistent calculation and save the obtained electronic structure in bxsf format [26] with the c2x code. [27] Ad hoc routines were developed to extract the EPC parameters as in eqs. (1) and (2). Finally, since we consider states and transitions in the whole 3D electronic structure, a DOS-weighted average of the computed EPC across the $\boldsymbol{q}$-points for each branch $\nu$ was performed to extract band-index specific EPC parameters to be used throughout the whole BZ:

$$D_{n,m,\nu} = \frac{\sum_q D_{\boldsymbol{q},\nu}^{n,m} DOS_{\boldsymbol{q},\nu}}{\sum_q DOS_{\boldsymbol{q},\nu}} \quad (3).$$

The selection of the vibrational modes to be considered in the protocol relies on the phonon DOS specifications and consists in selecting the modes bundled in the DOS. For pentacene, this encompasses the 20 modes of lowest frequency. The comprehensive EPC parametrized in Eq. (3) are regarded as the proper deformation potentials for inelastic processes involving the so-called non-polar optical phonons in the mobility calculation, as detailed in the next sub-section.

### c. Mobility

Crystalline OSCs exhibit a well-defined crystal structure, which results in a distinctive vibrational fingerprint. Adopting a description based on the electronic dispersions in the BZ of the reciprocal lattice, the mobility $\mu$ is evaluated from the conductivity $\sigma$ as:

$$\mu_{ij(E_F,T)} = \frac{\sigma_{ij(E_F,T)}}{n \cdot q_0} \quad (4)$$

where $i$ and $j$ are the Cartesian components $x$, $y$, and $z$, of the mobility and conductivity tensors, $E_F$ is the Fermi level, $T$ the temperature, $n$ the carrier density and $q_0$ the electronic charge. The conductivity is computed in the context of the linearized BTE as:[28]

$$\sigma_{ij(E_F,T)} = q_0^2 \int_E \Xi_{ij}(E) \left(-\frac{\partial f_0}{\partial E}\right) dE \quad (8)$$

The integrand of Eq. (8) contains the Transport Distribution Function (TDF) $\Xi_{ij}$ and the energy derivative of the equilibrium Fermi-Dirac distribution $f_0$. The TDF is defined as:

$$\Xi_{ij}(E) = \frac{2}{(2\pi)^3} \sum_n \sum_{\boldsymbol{k}_{n,E}} v_{i,\boldsymbol{k}_{n,E}} v_{j,\boldsymbol{k}_{n,E}} \tau_{i,\boldsymbol{k}_{n,E}}\, g_{\boldsymbol{k}_{n,E}} \quad (9)$$

In Eq. (9) $v$ is the band velocity, $\tau$ the relaxation time, and $g$ the electronic DOS. All these quantities are specific of each individual transport state $\boldsymbol{k}_{n,E}$, where $\boldsymbol{k}$ is the wave-vector, $n$ is the band index and $E$ its energy. So, the sum runs over all the transport states identified by their wave-vector $\boldsymbol{k}$, belonging to all the bands, having a certain energy. The DOS, $g_{\boldsymbol{k}_{n,E}}$ , is defined as $\frac{dA_{\boldsymbol{k}_{n,E}}}{|\vec{v}_{\boldsymbol{k}_{n,E}}|}$, where $dA_{\boldsymbol{k}_{n,E}}$ corresponds to the area of the surface element of the constant energy surface to which the $\boldsymbol{k}_{n,E}$ state belongs, associated to each specific $\boldsymbol{k}_{n,E}$ state. Therefore, the sum in Eq. 6 is performed for each constant energy surface to compute the energy dependent TDF, which will be then integrated as in Eq. (8). In this work, the tetrahedron method has been employed to construct the constant energy surfaces and extract the related quantities. [29,30]

The relaxation time of the state $\boldsymbol{k}_{n,E}$ for the transport along the direction $i$, associated with scattering by a phonon belonging to the branch $\nu$, is assessed from the inelastic scattering with non-polar optical phonons as (we omit the index $\nu$ for clarity): [24,25,30–32]

$$\frac{1}{\tau_{i,\boldsymbol{k}_{n,E}}} = \frac{1}{(2\pi)^3} \sum_{\boldsymbol{k}'_{n',E'}} \frac{\pi D^2_{n,n'}}{\rho \omega_0} \left( N_{\omega_0} + \frac{1}{2} \mp \frac{1}{2} \right) g_{\boldsymbol{k}'_{n',E'}} \left( 1 - \frac{v_{i,\boldsymbol{k}'_{n',E'}}}{v_{i,\boldsymbol{k}_{n,E}}} \right) \qquad (10)$$

where $D_{n,n'}$ is the deformation potential related to the electron-phonon scattering between the initial band $n$ and the final band $n'$, which includes intra- and inter-band processes, evaluated from Eq. (1) or (2), respectively; $\rho$ is the mass density. The effective frequency $\omega_{0,\nu}$ for the branch $\nu$ is evaluated from a DOS-weighted average over the selected portion of the phonon spectrum as $\omega_{0,\nu} = \frac{\sum_{\omega_\nu} \omega_\nu \cdot DOS(\omega_\nu)}{\sum_{\omega_\nu} DOS(\omega_\nu)}$; $N_{\omega_0}$ represents the phonon Bose-Einstein statistical distribution, while $g_{\boldsymbol{k}'_{n',E'}}$ is the DOS of the final state, belonging to the band $n'$ and at energy $E'$, which is either increased or decreased by $\hbar\omega$ for absorption or emission processes, respectively, denoted by "$-$" and "$+$" signs. The term $\left( 1 - \frac{v_{i,\boldsymbol{k}'_{n',E'}}}{v_{i,\boldsymbol{k}_{n,E}}} \right)$ approximates the momentum relaxation time, [31–33] which is the relevant type of relaxation time for computing transport coefficients. [25] Note that this definition of relaxation time is similar to what appears in other works on OSCs. [34] The calculation of the mobility is performed using the *ElecTra* simulator. [28,35]

Because the mobility tensor derived using equations is expressed in Cartesian coordinates, we project the electric field $\vec{\mathcal{E}}$ along the crystallographic axes. This is achieved by inverting the lattice vector matrix from the POSCAR VASP file:

$$\vec{\mathcal{E}_l} = \boldsymbol{l} * \mathrm{inv}(A) \qquad (11)$$

where $\boldsymbol{l}$ is the intended direction in the internal coordinates and $A$ is the lattice vector matrix. This allows us to express the electric fields along the internal cell axes, $\vec{\mathcal{E}_a}, \vec{\mathcal{E}_b}$, and $\vec{\mathcal{E}_c}$ in Cartesian coordinates. Next, from the conductivity tensor $\bar{\bar{\sigma}}$, given by Eq. (8), we compute a current density as:

$$\vec{J_l} = \bar{\bar{\sigma}} \vec{\mathcal{E}_l} \qquad (12).$$

Thus, we obtain a component of the conductivity tensor in internal coordinates:

$$\sigma_l = {|\overrightarrow{J_l}|}/{|\overrightarrow{\mathcal{E}_l}|} \tag{13}$$

and the related mobility:

$$\mu_l = {\sigma_l}/{n \cdot q_0} \tag{14}$$

Using this approach, we can link the computed mobility tensor in Cartesian coordinates to the mobility measured along specific crystal directions.

To perform the charge transport calculations, an additional non-self-consistent calculation on a finer mesh is needed. In this study, we adopted $\boldsymbol{k}$-samplings of $28 \times 21 \times 11$, $24 \times 18 \times 9$, and $28 \times 21 \times 11$, with cutoff energies of 800, 900, and 800 eV, for the LT, HT and TF polymorphs, respectively. Importantly, the energy resolution used to construct the constant energy surfaces is set at 1 meV, meaning that each surface is calculated every 1 meV for each band, from the band edge up to ~ 0.4 eV. Such a fine resolution proved to be essential for the treatment of the flatter bands of the OSC compared to inorganic compounds. [30,36–38] Due to the large bandgap of pentacene, we performed unipolar calculations separately for electron mobility in the conduction band (CB) and hole mobility in the valence band (VB).

## III. Results and Discussion

### a. Inherent Structures

Starting from the structures in the CCDC database,[39] from which we obtained HT,[40–42] LT, [40,41] and TF[43] CIF files, we first relax the atomic positions. The results are reported in Table 1. At the experimental lattice parameters of the polymorphs, the energy ranking is primarily governed by the VdW terms, demonstrating the importance of the dispersion contributions in stabilizing the structures. Next, we relaxed the cell parameters, to determine the inherent structures of the three polymorphs. The inherent structure of a system in a given configuration corresponds to the structure at mechanical equilibrium, found at the local minimum of the many-body potential energy hypersurface.[44] This minimum is reached by steepest descent minimization starting from the initial configuration. The unique local minima obtained from a set of experimental structures define the ‘‘natural’’ or ‘‘inherent’’ polymorphs the system can adopt. The approach effectively removes the noise due to the thermal expansion and is a reliable method for discerning experimental structures which may appear very similar [8,10]. The energy values reported in Table 2 for the three calculated local minima are genuinely distinct, confirming that the three experimental structures do correspond to different polymorphs. Moreover, the energy ranking is not determined by a single contribution but both electronic and dispersive interactions are relevant. As expected, the HT polymorph is the highest in potential energy, being in enantiotropic relationship with the LT form, with a transition temperature of 463K. [41] InterestinglyInterestingly Interestingly , the fact that the TF polymorph corresponds to a local minimum identifies it as a genuine structure the existence of which is not induced by the presence of a substrate, despite having never been observed in the bulk. Spectroscopically very similar to HT[12], TF is closer in energy to the LT polymorph. The genuinely different character of the three polymorphs is also signaled by the different Raman spectra, which is reported in Fig. S1 and further described in section **III.b**.

Table 1: total ($E_{tot}$), electronic ($E_{electr}$) and VdW ($E_{disp}$) energies, in eV, after relaxation of the atomic position at fixed cell.

| Atomic relax. | $E_{tot}$ | $E_{electr}$ | $E_{disp}$ |
|---|---|---|---|
| HT | -505.95925 | -499.89885 | -6.06040 |
| TF | -506.01846 | -499.79581 | -6.22265 |
| LT | -506.04048 | -499.70373 | -6.33675 |

Table 2: total ($E_{tot}$), electronic ($E_{electr}$) and VdW ($E_{disp}$) energies, in eV, after unit cell relaxation.

| u.c. relax. | $E_{tot}$ | $E_{electr}$ | $E_{disp}$ |
|---|---|---|---|
| HT | -506.05713 | -499.47794 | -6.57919 |
| TF | -506.06523 | -499.52383 | -6.54140 |
| LT | -506.06801 | -499.46723 | -6.60078 |

### b. Vibrational properties

The vibrational properties of the three polymorphs were computed to gain insight about their structural dynamics. We begin by observing that the spectral patterns are clearly distinct, confirming that each describes the dynamics of a different local minimum, see Fig. S1.

The Raman spectra computed at the experimental units cells, adjusted for laser frequency and temperature, are displayed in Fig. 1. The spectra are drawn as Lorentzian bands with a FWHM of 1/3 of the mean distance between the frequencies, in order to conform to the experimental features.[11] The peaks positions agree with the experiments within a few $cm^{-1}$, thus confirming the validity of our description in terms of pseudopotentials and electronic structure. [13,15,45,46] The computed relative intensities are brought to satisfactorily agree with the experiments once laser excitation frequencies and measurement temperature are accounted for. This is obtained with the formula $I = I_0 \frac{\nu}{(\nu-\nu_0)^4}\left(1 - \exp(\frac{-h\nu}{kT_0})\right)$, where $I_0$ , $\nu_0$, $T_0$ , and $\nu$ are the measured intensity, excitation frequency, temperature and vibration frequency, respectively, $h$ is the Planck constant and $k$ the Boltzmann constant. $I$ is the adjusted experimental intensity.[47] Calculated and experimental wavenumbers are reported in Table 3. To account for the frequency shifts observed for the lattice phonons computed at the experimental HT unit cell determined at 478 K by Siegrist et al.,[41] vibrational modes have also been calculated for the structure reported by Campbell[42]. While an approximately 2% increase of the cell volume in the high temperature structure has no effect on the intramolecular modes, a softening of the order of a few wavenumbers characterises the lattice phonons. Symmetry assignment was performed using the simmetry.py code, a post-processing routine which employs the projectors method[48]. In the triclinic $P_{\bar{1}}$ symmetry group, only modes of $A_g$ symmetry are Raman active. The low frequency modes are almost completely intermolecular in character and those of $A_g$ symmetry correspond to librations about the molecular inertia axis, while those of $A_u$ symmetry are translations.

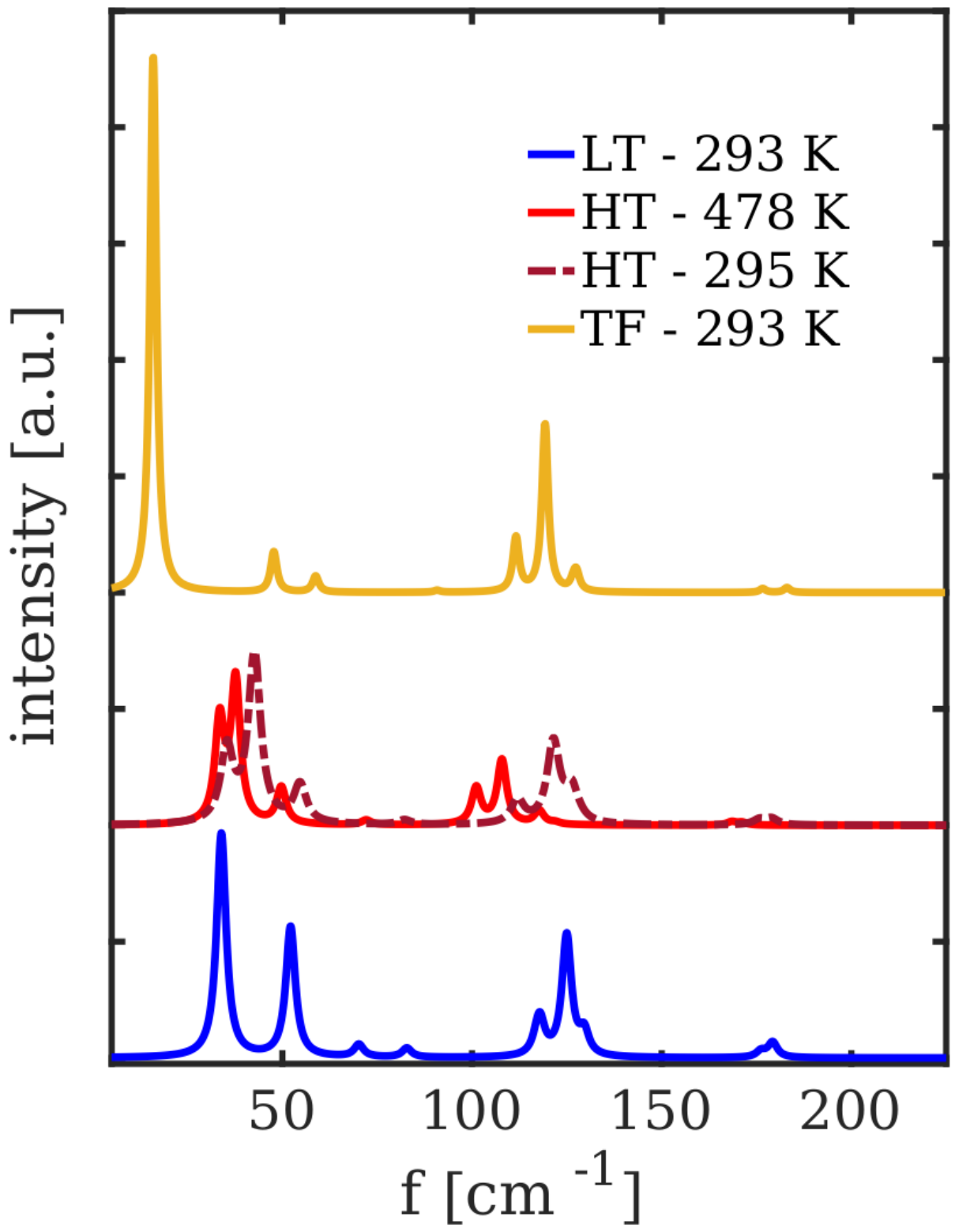

Figure 1: Calculated Raman spectra for the experimentally measured unit cells. The reported intensity is adjusted for the laser frequency and measurement temperature.

Table 3: Experimental Raman frequency for HT,[49,50] LT,[49,50] and TF,[12,49] compared with the computed values; we also report the polymorph nomenclature used in other works [49] and the name of the main author whom the polymorph structure is related to, and the temperature at which the spectra or the unit cell (u.c.) were measured. We show both gerade and ungerade modes for the sake of completeness. We restrict to the modes within the first bundle observable in the DOS. When computing the Raman frequency, we keep the experimental u.c. measured at the indicated temperature, thus a softening is expected for structures measured at high temperature.

| Experimental Polym. C (Campbell) 80 K | 300 K | Computed HT (Siegrist) u.c. @ 478 K | Computed HT (Campbell) u.c. @ 295 K | Experimental Polym. H (Holmes) 80 K | 300 K | Computed LT uc @ 293 K | Experimental TF 300 K | Computed TF u.c. @ 293 K |
|---|---|---|---|---|---|---|---|---|
| | | 25.6 ($A_u$) | 28.7 ($A_u$) | 44.9 | 33.1 | 33.9 ($A_g$) | | 16.0 ($A_g$) |
| 49.4 | 36.4 | 33.5 ($A_g$) | 35.4 ($A_g$) | | | 36.8 ($A_u$) | | 36.0 ($A_u$) |
| 54.9 | 45.5 | 37.7 ($A_g$) | 42.6 ($A_g$) | 65.5 | 52.2 | 52.2 ($A_g$) | 43.2 | 47.7 ($A_g$) |
| | | 49.6 ($A_u$) | 52.9 ($A_u$) | | | 54.0 ($A_u$) | | 48.0 ($A_u$) |
| 66.9 | 55.7 | 49.7 ($A_g$) | 54.6 ($A_g$) | | | 66.2 ($A_u$) | | 80.9 ($A_u$) |

| | | | | | | |
|---|---|---|---|---|---|---|
| | 58.3 ($A_u$) | 64.9 ($A_u$) | 84.3 69.8 | 70.2 ($A_g$) | 70* | 90.8 ($A_g$) |
| | 72.1 ($A_g$) | 82.2 ($A_g$) | 99.1 88.1 | 82.8 ($A_g$) | | 92.3 ($A_u$) |
| | 78.1 ($A_u$) | 87.6 ($A_u$) | | 85.4 ($A_u$) | | 111.6 ($A_g$) |
| | 82.2 ($A_u$) | 94.13 ($A_u$) | | 95.6 ($A_u$) | | 119.3 ($A_g$) |
| 99.8 94.4 | 101.1 ($A_g$) | 112.0 ($A_g$) | 132.2<br>122.5 | 117.1 ($A_g$) | | 120.4 ($A_u$) |
| 126.7 115.1 | 107.9 ($A_g$) | 121.5 ($A_g$) | 136.2<br>133.2 | 118.0 ($A_g$) | | 124.6 ($A_u$) |
| | 114.0 ($A_u$) | 122.1 ($A_u$) | | 121.0 ($A_u$) | 122.3 | 126.8 ($A_g$) |
| 140.6 132.3 | 117.7 ($A_g$) | 126.2 ($A_g$) | | 122.6 ($A_u$) | 132.3 | 127.5 ($A_g$) |
| | 119.2 ($A_u$) | 124.6 ($A_u$) | | 124.7 ($A_u$) | | 134.8 ($A_u$) |
| 147.5 | 121.9 ($A_g$) | 128.3 ($A_g$) | 144.1<br>149.5 | 124.9 ($A_g$) | | 137.1 ($A_u$) |
| | 123.0 ($A_u$) | 127.1 ($A_u$) | | | | |

* at $T$= 10 K.

Considering the key role played by dynamics and disorder in shaping the charge transport characteristics of the pentacene polymorphs, and the relevance of Raman spectroscopy as a probe of the processes involved, it is worth analysing in more detail the nature of the vibrational modes in the low frequency region of the spectrum. Simulations confirm the characteristic patterns detected in the experiments, with HT and TF spectra more similar to each other than to LT.[12] The eigenvector analysis reveals that the Raman modes below 120-130 $cm^{-1}$ are predominantly of intermolecular nature. As expected, the lowest energy vibrations involve mixed librations around the normal-to-plane (N) and short-in-plane (M) molecular inertia axes, which correspond to the highest moments of inertia. At higher energy, the modes correspond to almost pure rotations around the long-in-plane (L) molecular axis. The highest-energy mode, however, exhibits a mixed intramolecular character, particularly in the HT and TF phases. It should be noticed that the calculations reproduce the experimental evidence that the high frequency pattern is very similar for the three polymorphs, whereas major differences affect the lowest frequency range, where in fact, the TF Raman spectrum has never been clearly detected [12,49] because in thin films HT and TF grow concomitantly and their appearance is found to be regulated by the film's growth rate.[12] As observed previously,[12] such a feature must correspond to a characteristic of the pentacene polymorph structures, and finds its explanation upon the analysis of the phonon dispersion curves.

Table 4: calculated wavenumbers of the lowest lattice phonon modes of HT, LT and TF polymorphs at Γ (with symmetry) with squared translational (T) and rotational (R) components along and around the three main inertia axes (L, M, N), L and M in the molecule plane, N normal to the molecular plane. The modes in bold are the lattice phonons well-visible in the Raman spectrum, the modes in bold and italic, dark grey, are mostly internal modes well-observed in the Raman spectrum, the modes not in bold are lattice (black) or internal (faint gray) modes which are poorly or not observed.

| Freq. $cm^{-1}$ | Symmetry | $T_L$ | $T_M$ | $T_N$ | $R_L$ | $R_M$ | $R_N$ |
|---|---|---|---|---|---|---|---|
| **HT** – Siegrist | | | | | | | |
| **33.5** | $\mathbf{A_g}$ | **0** | **0** | **0** | **1** | **14** | **85** |
| **37.7** | $\mathbf{A_g}$ | **0** | **0** | **0** | **0** | **43** | **54** |
| **49.7** | $\mathbf{A_g}$ | **0** | **0** | **0** | **1** | **55** | **43** |
| 72.1 | $A_g$ | 0 | 0 | 0 | 0 | 83 | 16 |
| **101.1** | $\mathbf{A_g}$ | **0** | **0** | **0** | **88** | **0** | **0** |
| **107.9** | $\mathbf{A_g}$ | **0** | **0** | **0** | **95** | **0** | **1** |
| ***117.7*** | $\boldsymbol{A_g}$ | ***0*** | ***0*** | ***0*** | ***14*** | ***3*** | ***0*** |
| 121.9 | $A_g$ | 0 | 0 | 0 | 0 | 0 | 1 |
| **LT** – Holmes | | | | | | | |
| **33.9** | $\mathbf{A_g}$ | **0** | **0** | **0** | **0** | **61** | **38** |
| **52.2** | $\mathbf{A_g}$ | **0** | **0** | **0** | **0** | **12** | **86** |
| **70.2** | $\mathbf{A_g}$ | **0** | **0** | **0** | **1** | **30** | **65** |
| **82.8** | $\mathbf{A_g}$ | **0** | **0** | **0** | **1** | **90** | **6** |
| ***117.1*** | $\boldsymbol{A_g}$ | ***0*** | ***0*** | ***0*** | ***4*** | ***3*** | ***0*** |
| **118.0** | $\mathbf{A_g}$ | **0** | **0** | **0** | **90** | **0** | **1** |
| **125.0** | $\mathbf{A_g}$ | **0** | **0** | **0** | **82** | **1** | **1** |
| ***129.6*** | $\boldsymbol{A_g}$ | ***0*** | ***0*** | ***0*** | ***19*** | ***2*** | ***2*** |
| **TF** | | | | | | | |
| **16.0** | $\mathbf{A_g}$ | **0** | **0** | **0** | **0** | **14** | **85** |
| **47.7** | $\mathbf{A_g}$ | **0** | **0** | **0** | **0** | **35** | **58** |
| **58.8** | $\mathbf{A_g}$ | **0** | **0** | **0** | **0** | **56** | **41** |
| 90.8 | $A_g$ | 0 | 0 | 0 | 0 | 85 | 14 |
| **111.6** | $\mathbf{A_g}$ | **0** | **0** | **0** | **90** | **0** | **0** |
| **119.3** | **$A_g$** | **0** | **0** | **0** | **96** | **0** | **0** |
| *126.8* | $A_g$ | *0* | *0* | *0* | *8* | *1* | *0* |
| *127.5* | $A_g$ | *0* | *0* | *0* | *5* | *8* | *1* |

Phonon dispersions and corresponding DOS are reported in Fig. 2. The dispersions along directions connecting high symmetry $\boldsymbol{q}$-points are shown in Fig.s 2a-c. It is important to note that the first 20 modes are bundled together throughout the BZ. This feature is highlighted in Fig. 2d, where the DOS of the polymorphs are reported. Thus, we consider them as 20 scattering channels, following the common approach in semiconductors,[25] and, due to the dispersion through the BZ, we assign to each of them an effective phonon frequency $\omega_{0,\nu}$ .[11] Because of their bundled character, we consider all of them and cut our region of interest at the first drop to zero of the phonon DOS.

Vibrational spectroscopy provides lattice mode frequencies at the Γ point of the BZ, where all unit cells move in phase. The phonon dispersion curves, which describe the dependence of vibrational frequencies on the wavevector, offer insight into intermolecular interactions among different crystal cells. Analysis of the curves for the three pentacene polymorphs in Fig. 2 reveals that all modes of intermolecular character exhibit significant dispersion along the *a**b** directions (approximately in the *ab* crystal plane), demonstrating the strength of ab intra-layer interactions. In contrast, only the lowest-frequency modes, identified in the Raman spectrum as librations that tilt molecules, are dispersed along *c**, the direction nearly perpendicular to the *ab* plane, where interlayer interactions are present. Librations about the L axis remain dispersionless in this direction, indicating their insensitivity to interlayer interactions, consistently with the fact that these phonons have the same description in the three polymorphs. This computational finding reinforces the results of lattice dynamics calculations based on empirical interatomic potentials for the TF structure and extends them to all pentacene polymorphs.

The occurrence of structural disorder in the *c** direction disrupts the periodical interaction potential along that axis, thus preventing the establishment of a correlation length for collective phonon motions, that is the distance over which the atomic displacements of the vibrations remain correlated. This mechanism has been invoked to explain both the vanishing intensities and the band broadenings affecting the lowest frequencies lattice phonons observed in the films of pentacene polymorphs TF and HT,[12] where the degree of the layer-by-layer order along the *c* axis depends on the deposition rate of the material. In contrast, the persistence of the high frequency phonons indicates that the order on the *ab* plane, parallel to the film substrate, is preserved. In the actual samples, such a disorder is expected to influence the EPC along the *c** direction in a number of ways. If the phonon mode cannot propagate coherently due to structural disorder, then the electron interaction would be less effective. We further elaborate on this point in the EPC section.

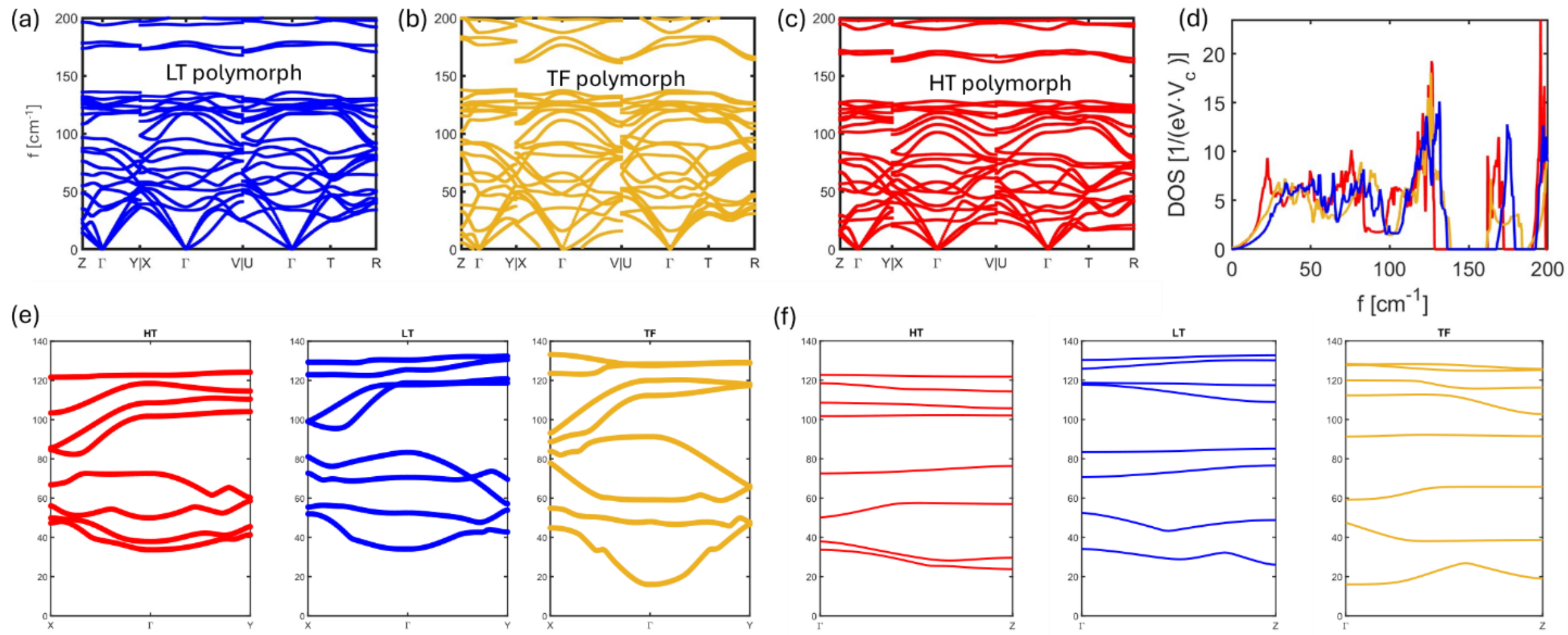

Figure 2: Phonons dispersions for the three polymorphs (a-c) and their DOS (d). In (e) and (f), we compare the dispersions of the Raman active modes of the three polymorphs, in the *ab* plane and along the *c* axis, respectively.

### c. Electronic structure

The electronic structure of pentacene has been investigated in detail elsewhere.[51] We focus here primarily on the comparison of the electronic DOS of the three polymorphs, plotted in Fig. 3. Solid lines are used for the topmost VB, while dash-dot lines denote the second one. The overlap region is shadowed, and the $B_W$ of the topmost VB is report in the legend. For the LT polymorph, the computed $B_W$ compares well with the Angle-Resolved PhotoEmission Spectroscopy (ARPES) measurements of 206 meV (T = 110 K) and 175 meV (T = 75 K).[52,53] ARPES measurements, presumably on LT phase samples, reveal bands narrower than what we obtain from DFT. However, we note that the larger bandwidth that we observe derives from the X-Y and T-U directions, which spans the $(0,0,k_z)$ planes with $k_z$ = 0 and 0.5, respectively, according to the Bilbao Crystallographic Data Center nomenclature,[19,20] thus following a line not passing from Γ. Since ARPES measurements are done along the reciprocal lattice direction from Γ to $(1\bar{1}0)$, which correponds to Γ-V in our plot, a smaller detected bandwidth by ARPES is expected.

Furthermore, in the DOS plot in Fig. 3, we highlight the overlap region between the two bands by shadowing it. This is the energy range where the inter-band scattering, Eq. (2), plays a role. We see that it is negligible for the HT phase, because of the substantial absence of overlap between the DOS, so that the fraction of carriers which can pass from one band to the other is negligible. The inter-band processes play a larger, yet minor, role in the LT because the DOS overlap involves higher energy carrier. Differently, it is expected to be relevant in the TF polymorph, where it starts around 100 meV from the edge. Moreover, the EPC for the inter-band processes is seen as a vibrational modulation of the interaction between the two non-equivalent molecules inside the unit cell. Since $B_W$ is comparable to the bands separation, the interaction between the molecules inside the cell is similar with the one between molecules of other cells, i.e. we are not allowed to think in terms of dimer. Thus, we expect the inter-band EPC is stronger also out of Γ, as we confirm in Fig. S3.

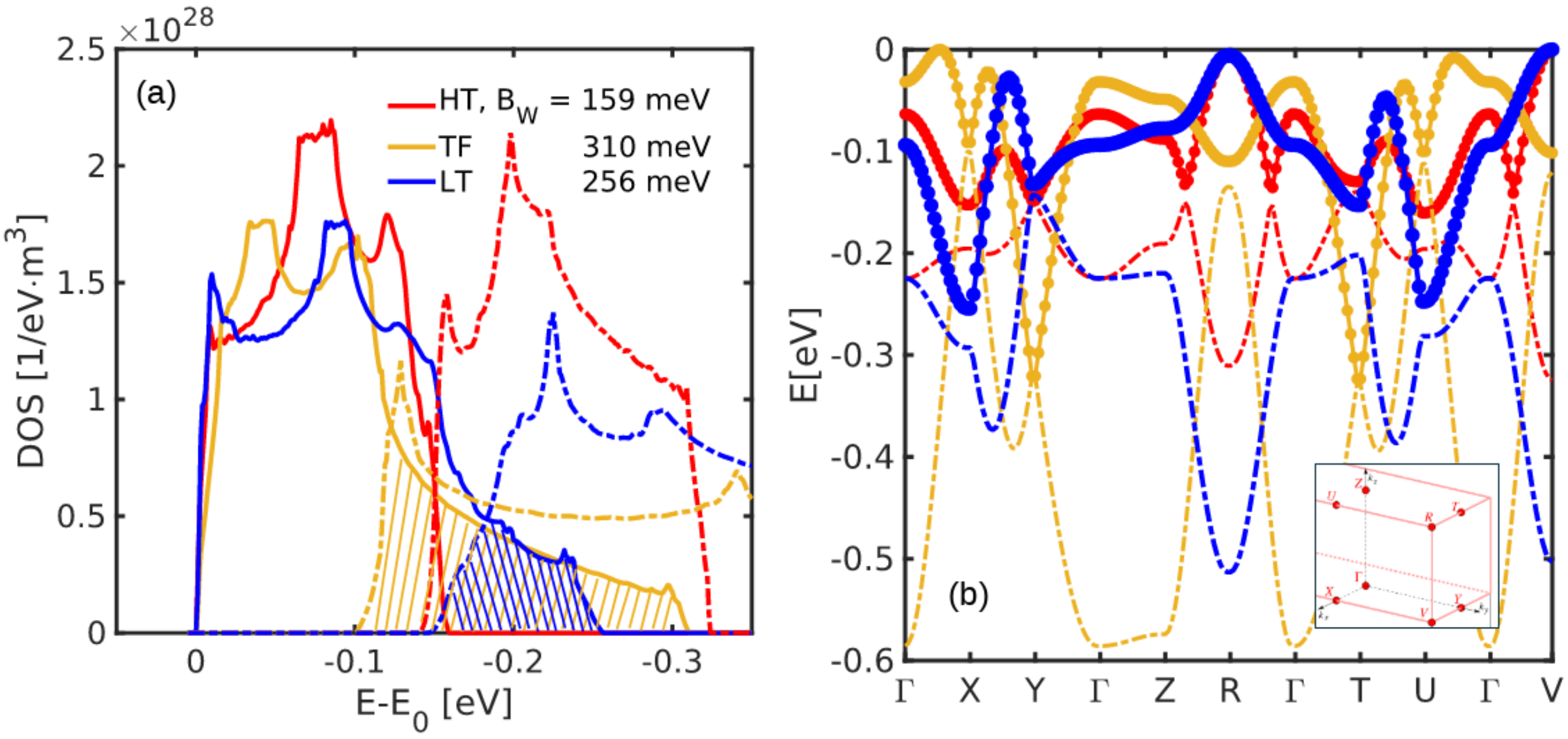

Figure 3: VB electronic structure of the three pentacene polymorphs, HT (red) TF (dark yellow) and LT (blue): comprehensive DOS with indication of the bandwidth of the uppermost VB, and projections along certain directions; solid lines are for the uppermost band and dash-dot lines for the lower one. The inset in (b) reports the BZ wedge with the high symmetry points as taken from the Bilbao Crystallographic Data Center nomenclature[19,20].

### d. EPC parameters

In this section, we present the EPCs evaluated according to Eq. s (1)-(3). They have been computed with the Phonopy code on unit cells with atoms displaced along each eigenmode, following convergence tests that identified 0.0025 Å as optimal displacement value, the results of the convergence tests are reported in Fig. S2. In Table 5, we report the mode and the **q**-point pair $(\boldsymbol{q}, \nu)$ for which the EPC exceeds 85 % of the highest value. We first observe that, in Γ, there are no modes that significantly contribute to the EPC. This demonstrates the importance of not confining the analysis to the BZ center,[54] even though only Γ modes are spectroscopically accessible. Also, we notice that the HT polymorph has four modes with strong EPC, LT has only two modes, with EPC comparable to the HT, and TF has two modes with stronger EPC than both the other polymorphs. In Table 5, we also report the effective EPC for that mode and $\boldsymbol{q}$-point, evaluated as the EPC squared divided by the phonon energy, $\mathrm{EPC}^2/\hbar\omega$, after eq. (10). We see that the effective scattering strength in the LT drops while in the other polymorphs stays high; as we see in section **III.d**, this influences the mobility.

We notice that in all three polymorphs, the largest EPC values are obtained for $\boldsymbol{q}$-points which correspond to a direction with a *c* axis component, nominally U and R. The disorder effect described in the previous section, which loss of coherence, disrupts periodicity and the BZ concept loses sense along that direction. Hence, the EPC at R and U cannot be defined, with a decrease of the overall EPC, as per eq. (3). However, vibrational degrees of freedom are expected to appear as a Gaussian distribution of frequencies without dispersion relationship, i.e. lacking any definite $\boldsymbol{q}$-value dispersion relationship in phase and frequency. These modes may contribute to EPC, but their effect cannot be precisely quantified at this level of description. If this approximate of "surrogate" EPC might be lower than the values calculated at the R and U points, which is likely, given that those represent some of the strongest couplings, then the overall EPC would be reduced, potentially leading to an enhancement in carrier mobility. This can be the reason for the experimentally observed larger $\mu$ in films with a corresponding low inter-layer correlation, controlled by the growth rate, along $c^*$.[55]

In addition to the $(\boldsymbol{q},\nu)$ pair with the highest EPC values, Table 5 reports the contribution to the effective EPC strength $\mathrm{EPC}^2/\hbar\omega$ of the branch to which the pairs belong, following the calculation of the DOS-weighted average over the BZ as in eq. (3). It appears that, in HT and TF polymorphs, ~60% of the EPC strength can be ascribed to three modes, for the TF phase the mode #5 does not have particularly relevant individual $(\boldsymbol{q},\nu)$ pairs but the overall branch accounts for a generally large EPC. Differently, in the LT polymorph the 60% of the EPC strength is due to a single mode, with the largest contribution coming from zone boundaries.

The largest EPC occurs away from the Γ-point in all polymorphs, suggesting that spectroscopic methods relying solely on measurements at Γ [6,56,57] may provide an incomplete picture of the coupling landscape.

Table 5: $\boldsymbol{q}$-points and phonon branch numbers featuring the highest EPC values in the three pentacene polymorphs. The effective EPC as per eq. (10) is reported as well, together with the mode symmetry and decomposition along the main molecular inertia axes. The contribution of the whole branch after the DOS weighted average over the BZ is also given. The $\boldsymbol{q}$-points in bold have a component along the $z$ direction.

| Polymorph | $\boldsymbol{q}$-point | Freq. $cm^{-1}$ *Branch #* | EPC [eV/Å] | $EPC^2/\hbar\omega$ [eV/√Å] | sym. | $T_L$ | $T_M$ | $T_N$ | $R_L$ | $R_M$ | $R_N$ | mode contribution to EPC over BZ |
|---|---|---|---|---|---|---|---|---|---|---|---|---|
| | | | | | | | | | | | | |
| **HT** | V | 45.3 *# 5* | 0.35 | 21.8 | A | 5 | 34 | 8 | 5 | 2 | 10 | 25 % |
| **HT** | **R** | 41.6 *# 4* | 0.33 | 21.1 | A | 0 | 3 | 1 | 1 | 14 | 20 | 16 % |
| **HT** | X | 46.9 *# 5* | 0.30 | 15.5 | A | 0 | 44 | 9 | 12 | 0 | 10 | 25 % |
| **HT** | **U** | 49.9 *# 6* | 0.30 | 14.5 | A | 0 | 31 | 1 | 6 | 4 | 20 | 16 % |
| | | | | | | | | | | | | |
| **TF** | V | 47.9 *# 6* | 0.49 | 40.4 | A | 1 | 2 | 1 | 3 | 4 | 67 | 23 % |
| **TF** | **U** | 45.7 *# 3* | 0.43 | 32.6 | A | 4 | 26 | 1 | 2 | 1 | 16 | 20 % |
| **TF** | Whole BZ | *# 5* | | | | | | | | | | 21 % |
| | | | | | | | | | | | | |
| **LT** | **U** | 26.7 *# 2* | 0.27 | 22.0 | A | 67 | 0 | 1 | 1 | 0 | 3 | 60 % |
| **LT** | X | 39.2 *# 2* | 0.24 | 11.9 | A | 46 | 2 | 1 | 0 | 2 | 4 | 60 % |

### e. Mobility

In this section, we summarize the results achieved in terms of transport properties. First, we observe that the phonon-limited study reproduces the experimental rank. We use the approach in eq.s (11) to (14) to obtain the mobility projected along arbitrary 3D orientations and compute an average mobility $\mu_r$ using more than $10^6$ random orientations. For the LT, HT and TF, at room temperature, we find $\mu_r$ = 28.8, 6.5, and 5.7 $cm^2/Vs$, respectively. To enforce our comparison with experimental results, in Table 6, we compare the computed mobilities with the experimental values of the best-case scenarios we found in literature for bulk single-crystal (s.c.), [58] single-crystal field effect transistor (s.c.-FET),[59] and thin film field effect transistor (TF-FET).[60] The calculated mobility is along $b$ axis, $\mu_b$, in the $ab$ plane, $\mu_{<ab>}$ , and also averaged over two million of random orientations, $\mu_r$.

In Table 6, for the TF-FET case, we report also the case of a mixture of two polymorphs, where the mixture is treated as inhomogeneous two-components medium,[61] leading to:

$$\mu = \mu_1^{f_1} \mu_2^{1-f_1} \quad (15),$$

Where 1 and 2 are the two phases and $f_1$ is the volume fraction of the first phase. In our estimates in Table 6, we assumed that TF and HT coexist [12] in equal amount ($f_1 = 0.5$). Also, the different islands can have different structures at the coalescence boundary, with an impact on the resulting mobility.[55] Thus, we phenomenologically introduce a scattering term with characteristic energy equal to the energy distance between the top-valence band and other satellite valleys at lower energy in the HT phase, $\Delta E_{\mathrm{V}}^{\mathrm{HT}}$, which is 32.5 meV. We consider only HT since TF appears not to have such satellite valleys but only a valence band maximum without relative maxima. We treat such boundary scattering on the foot of the Fermi golden rule, using such characteristic energy as scattering potential $U_b$ and the energy dependent density of the states of the unit cell $g_{uc(E)}$. Thus, we have a scattering time $\tau_{b(E)}$ for this boundary scattering:

$$\frac{1}{\tau_{b(E)}} = \frac{2\pi}{\hbar} (U_b)^2 g_{uc(E)} \quad (16).$$

We used eq. (16) also to account, in a macroscopic and phenomenological way, of structural defects in single crystals, where an effective barrier of ~ 40 meV brings the phonon mobility in the measured range. Lastly, after the Kelvin probe microscopy (KPM) measurements on in operando pentacene FET, which revealed a ~ 0.1 V voltage drop at any coalescence boundary between islands, [3] we used eq. (16) with $U_b = 0.1$ eV, together with a 1:1 polymorph mixture. The obtained $\mu$ is lower than the value measured in FET with polymeric gate,[60] but is closer to what measured in the same experiment of KPM measurements,[3] which featured oxide gate dielectrics. Thus, our preliminary results suggest that the strong impact of the dielectric/OSC interface [62] occurs also at the level of crystal quality and defectivity of the OSC.

Table 6: comparison between experimental and calculated $\mu$ values, the crystal direction is at subscript and considered polymorph phase is in parenthesis. When eq. (16) is included, it is indicated. The asterisk indicates a mixture treated as from eq. (15).

| System | Experimental $\mu$ [$cm^2$/Vs] | Calculated $\mu$ [$cm^2$/Vs] |
|---|---|---|
| s.c. | 35 [58] | $\mu_b$ = 52 (LT) |
| s.c. - FET | 5.7 [59] | $\mu_{\langle ab \rangle}$ = 36 (LT) 5.9 (LT, $U_b = 40\ meV$) |
| TF-FET | 3[60] | $\mu_{<ab>}$ = 7.1 (TF)<br>$\mu_{r}$ = 5.7 (TF)<br>$\mu_{<ab>*}$ = 4.2 (TF:HT=1:1, $U_b = \Delta E_{\mathrm{V}}^{\mathrm{HT}} = 32.5\ meV$)<br>$\mu_{<ab>*}$ = 1.6 (TF:HT=1:1, $U_b = 0.1\ eV$) |

## IV. Conclusions

We employed a robust first-principles approach to investigate why closely related polymorphic systems, such as the three known crystal forms of pentacene, may in fact display markedly different charge mobilities. Our analysis of non-local electron–phonon couplings over the full BZ reveals that even structurally similar polymorphs can be characterized by distinct “killer” phonon modes.[6] Thus, we challenge the notion of killer phonon, i.e. the existence of a dominant vibrational mode. Instead, our studies reveal that the mobility is limited by more modes with diverse wave-vector, responsible for the scattering of the charge carriers and limiting the mobility. Importantly, these modes are not restricted to the BZ center detected by the spectroscopic methods (i.e., they possess non-zero wavevectors). Even in the LT polymorph, where 60% of the effective EPC strength is ascribed to one phonon branch only, there are two dominant phonons belonging to the same branch but different wave-vector. Since the identity and character of these detrimental phonons appear to depend even on structural details in closely related forms, their identification remains a complex task, which requires a rigorous theoretical framework. What we can deduce at present is that in the system with larger mobility, the stronger EPC is for modes involving molecules translations along the long inertia molecular axis. However, at which extent this conclusion is general and how to derive molecule design indications are still open tasks.

Our phonon-limited mobility calculations accurately reproduce the experimental trend in the three pentacene polymorphs, with average mobility values of 28.8, 6.5, and 5.7 cm²/V·s for LT, HT, and TF phases, respectively. These values are consistent with the highest experimental mobility recorded for single crystals, FETs, and thin films. Importantly, we also modeled mixed-phase systems and incorporated additional scattering mechanisms to account for boundary effects and structural defects by means of a phenomenological treatment based on Fermi's golden rule. These results highlight how subtle structural differences, phase mixing, and interface effects can dramatically influence charge mobility in real devices.

Our analysis also sheds light on how a disorder along the $c^*$ can lead to an effective phonon confinement in a quasi 2D system, because of the suppression of the phonons with wave-vector $\boldsymbol{q}$ oriented out of *ab* plane. This effect removes some electron–phonon scattering processes leading to a decreased EPC, assuming that structural order is preserved on the *ab* plane and interface-related scattering is minimal and suggests the reason for enhanced charge transport in thin films and monolayers.

**Acknowledgement**

We are thankful to prof. Raffaele Guido Della Valle for support and in-depth analysis about the phonons symmetry, and for insightful discussions on the presented results, and to prof. Alberto Girlando for fruitful discussions. We acknowledge the CINECA award under the ISCRA initiative, for the availability of high-performance computing resources and support. We acknowledge funding from the European Union–Next-Generation EU via the Italian call PRIN 2022, project code 2022XZ2ZM8, "POLYPHON". T.S. thanks the Programma per Giovani Ricercatori "Rita Levi Montalcini" year 2020 (grant PGR20QN52R) of the Italian Ministry of University and Research (MUR) for the financial support. We acknowledge funding for the VASP license from the company M.M.B. s.r.l., Faenza (RA) Italy.